\documentclass[11pt]{article}
\usepackage{moriond,epsfig}
\usepackage{amssymb,amsmath}
\def\be{\begin{equation}}
\def\ee{\end{equation}}
\def\bea{\begin{eqnarray}}
\def\eea{\end{eqnarray}}

\def\nbox#1#2{\vcenter{\hrule \hbox{\vrule height#2in
\kern#1in \vrule} \hrule}}
\def\sq{\,\raise.5pt\hbox{$\nbox{.09}{.09}$}\,}
\def\sqb{\,\raise.5pt\hbox{$\overline{\nbox{.09}{.09}}$}\,}
\newcommand{\bes}{\begin{subequations}}
\newcommand{\ees}{\end{subequations}}

\def\Real{\mathbb{R}}

\begin{document}
%\vspace*{4cm}
\title{Conformal Invariance, Dynamical Dark Energy and the CMB}

\author{Emil Mottola}

\address{Theoretical Div., Los Alamos National Laboratory,
Los Alamos, NM 87545 USA\\
E-mail: emil@lanl.gov\\ \vspace{1mm} and \vspace{1mm}\\
Theoretical Physics Group, PH-TH, CERN
CH-1211, Geneva 23, Switzerland\\
E-mail: emil.mottola@cern.ch}

\maketitle\abstracts{General Relativity receives quantum corrections 
relevant at cosmological distance scales from conformal scalar degrees of freedom 
required by the trace anomaly of the quantum stress tensor in curved space. In the theory
including the trace anomaly terms, the cosmological ``constant"  becomes dynamical 
and hence potentially dependent upon both space and time. The fluctuations of these 
anomaly scalars may also influence the spectrum and statistics of the Cosmic 
Microwave Background. Under the hypothesis that scale invariance should be promoted 
to full conformal invariance, an hypothesis supported by the exact equivalence of the 
conformal group of three dimensions with the de Sitter group $SO(4,1)$, the form of the 
CMB bispectrum can be fixed, and the trispectrum constrained. The non-Gaussianities
predicted by conformal invariance differ from those suggested by 
simple models of inflation.\vspace{3mm} \\
Preprint Numbers: LA-UR-10-04603,\ CERN-PH-TH 2011-039}

\section{Cosmological Dark Energy and the Effective Field Theory of Gravity}

Observations of type Ia supernovae at moderately large redshifts ($z\sim 0.5$ to $1$) 
have led to the conclusion that the Hubble expansion of the universe is {\it accelerating}.\cite{SNI} 
According to Einstein's equations this acceleration is possible if and only if the energy 
density $\rho$ and pressure $p$ of the dominant component of the universe 
satisfies the inequality, 
\be
\rho + 3 p \equiv \rho\  (1 + 3 w) < 0\,.
\label{accond}
\ee
A vacuum energy with $\rho_v > 0$ and $w\equiv p_v/\rho_v = -1$ leads to an 
accelerated expansion, a kind of ``repulsive" gravity in which the relativistic
effects of a negative pressure can overcome a positive energy density in
(\ref{accond}). Taken at face value, the observations imply that some $74\%$ of 
the energy in the universe is of this hitherto undetected $w=-1$ dark 
variety. This leads to a non-zero inferred cosmological term in Einstein's 
equations of 
\be
\Lambda_{\rm meas} \simeq (0.74)\, \frac{3 H_0^2}{c^2} 
\simeq 1.4 \times 10^{-56}\ {\rm cm}^{-2}
\simeq  3.6 \times 10^{-122}\ \frac{c^3}{\hbar G}\,.
\label{cosmeas}
\ee
Here $H_0$ is the present value of the Hubble parameter, approximately 
$73$ km/sec/Mpc  $\simeq 2.4 \times 10^{-18}\, {\rm sec}^{-1}$. The last 
number in (\ref{cosmeas}) expresses the value of the cosmological term
inferred from the SN Ia data in terms of Planck units, 
$L_{\rm pl}^{-2} = \frac{c^3}{\hbar G}$. Explaining the value of this 
smallest number in all of physics is the basic form of the {\it cosmological constant problem.}

If the universe were purely classical, $L_{pl}$  would vanish and $\Lambda$, 
like the overall size or total age of the universe, could take on any value 
whatsoever without any technical problem of naturalness. On the other hand, 
if $G=0$ and there are also no boundary effects to be concerned with, then 
the cutoff dependent zero point energy of flat space could simply be subtracted, 
with no observable consequences. A naturalness problem arises only when 
the effects of quantum zero point energy on the large scale curvature of spacetime 
are considered. Thus this is a problem of the gravitational energy of the quantum 
vacuum or ground state of the system at {\it macroscopic} distance scales, 
very much greater than $L_{pl}$, when both $\hbar \neq 0$ and $G \neq 0$.

The treatment of quantum effects at distances much larger than any ultraviolet 
cutoff is precisely the context in which effective field theory (EFT) techniques 
should be applicable. This means that we assume that we do not need to know 
every detail of physics at extremely short distance scales of $10^{-33}$ cm or 
even $10^{-13}$ cm in order to discuss cosmology at $10^{28}$ cm scales. 
In extending Einstein's classical theory to take account of the quantum properties
of matter, the classical stress-energy tensor of matter $T^a_{\ b}$ becomes a
quantum operator, with an expectation value $\langle T^a_{\ b}\rangle$. In this
{\it semi-classical} theory with both $\hbar$ and $G$ different from zero,
quantum zero-point and vacuum energy effects first appear, while the
spacetime geometry can still be treated classically. Since the expectation 
value $\langle T^a_{\ b}\rangle$ suffers from the quartic divergence, 
a regularization and renormalization procedure is necessary in order to define 
the semi-classical EFT. The result of the renormalization program for quantum fields 
and their vacuum energy in curved space is that General Relativity can be viewed 
as a low energy quantum EFT of gravity, provided that the classical Einstein-Hilbert 
classical action is augmented by the additional terms required by the {\it trace anomaly}
when $\hbar \neq 0$.

Massless quantum matter or radiation fields have stress-energy tensors which
are traceless classically. However it is impossible to maintain both conservation
and tracelessness of the quantum expectation value $\langle T^a_{\ b}\rangle$.
Instead a well-defined conformal or trace anomaly for this expectation value
in curved spacetime is obtained,\cite{BirDav} {\it i.e.}
\be
\langle T^a_{\ a} \rangle
= b F + b' \left(E - \frac{2}{3}\sq R\right) + b'' \sq R \,,
\label{tranomgen}
\ee
where
\bes\bea
&&E \equiv ^*\hskip-.2cmR_{abcd}\,^*\hskip-.1cm R^{abcd} = 
R_{abcd}R^{abcd}-4R_{ab}R^{ab} + R^2 
\,, \label{EFdef}\\
&&F \equiv C_{abcd}C^{abcd} =
 R_{abcd}R^{abcd} - 2 R_{ab}R^{ab}  + \frac{R^2}{3}\,.
\eea\ees
in terms of the Riemann curvature tensor $R_{abcd}$. 
The coefficients $b$ and $b'$ in (\ref{tranomgen}) do not depend on any ultraviolet 
short distance cutoff, but instead are determined only by the number and spin of 
massless fields via
\bes\bea
b &=& \frac{\hbar}{120 (4 \pi)^2}\, (N_S + 6 N_F + 12 N_V)\,,\\
b'&=& -\frac{\hbar}{360 (4 \pi)^2}\, (N_S + \frac{11}{2}N_F + 62 N_V)\,,
\label{bprime}
\eea \label{bbprime}\ees
\noindent with $(N_S, N_F, N_V)$ the number of massless fields of spin 
$(0, \frac{1}{2}, 1)$ respectively. The number of massless fields of each spin 
is a property of the low energy effective description of matter, having no direct 
connection with physics at the ultrashort Planck scale. Indeed massless fields 
fluctuate at all distance scales and do not decouple in the far infrared, relevant for
cosmology. 

One can find a covariant action functional whose variation gives the trace anomaly
(\ref{tranomgen}). This functional is {\it non-local} in terms of just the curvature, and 
hence describes long distance infrared phyiscs. The non-local anomaly action may be
put into a local form, but only by the introduction of two new scalar fields $\varphi$ and 
$\psi$. Then the local effective action of the anomaly in a general curved space may
be expressed in the form\cite{MotVau,NJP}  
\be
S_{anom} = b' S^{(E)}_{anom} + b S^{(F)}_{anom}\,,
\label{allanom}
\ee
where
\bea
&& \hspace{-5mm}S^{(E)}_{anom}\equiv \frac{1}{2}\int \!d^4x\sqrt{-g}\left\{
-\left(\sq \varphi\right)^2 + 2\left(R^{ab}- \frac{R}{3} g^{ab}\right)
(\nabla_a \varphi)(\nabla_b \varphi) + \left(\!E - \frac{2}{3} \sq R\right)
\varphi\right\}\,;\nonumber\\
&& S^{(F)}_{anom} \equiv \,\int\,d^4x\,\sqrt{-g}\ \left\{ -\left(\sq \varphi\right)
\left(\sq \psi\right) + 2\left(R^{ab} - \frac{R}{3}g^{ab}\right)(\nabla_{\mu} \varphi)
(\nabla_{\nu} \psi)\right.\nonumber\\
&& \qquad\qquad\qquad + \left.\frac{1}{2} C_{abcd}C^{abcd}
\varphi + \frac{1}{2} \left(E - \frac{2}{3} \sq R\right) \psi \right\}\,.
\label{SEF}
\eea
\vspace{-.4cm}

\noindent
The free variation of the local action (\ref{allanom})-(\ref{SEF}) with respect to $\psi$ 
and $\varphi$ yields their eqs. of motion. Each of these terms when varied with 
respect to the background metric gives a stress-energy tensor in terms of these 
anomaly scalar fields $\varphi$ and $\psi$. The scalar fields of the local form 
(\ref{SEF}) of the anomaly effective action describe massless scalar degrees of 
freedom of low energy gravity, not contained in classical General Relativity.  
The effective action of low energy gravity is thus
\be
S_{eff}[g] = S_{_{EH}}[g] + S_{anom}[g;\varphi,\psi]
\label{Seff}
\ee
with $S_{_{EH}}$ the Einstein-Hilbert action of classical General Relativity
and $S_{anom}$ the anomaly action given by (\ref{allanom})-(\ref{SEF}). 

\section{Dynamical Dark Energy}

In order to understand the {\it dynamical} effects of the kinetic terms in the 
anomaly effective action (\ref{SEF}), one can consider simplest case of the quantization
of the conformal factor in the case that the fiducial metric is flat, {\it i.e.} $g_{ab} = 
e^{2 \sigma}\eta_{ab}$. In this case the Wess-Zumino form of the effective anomaly action 
(\ref{allanom}) simplifies to\cite{NJP,AntMot}
\be
S_{anom}[\sigma] = -\frac{Q^2}{16\pi^2} \int d^4 x \ (\sq \sigma)^2 \,,
\label{flata}
\ee
where 
\be
Q^2 \equiv -32 \pi^2 b'\,.
\label{Qdef}
\ee
This action quadratic in $\sigma = \varphi/2$ is the action of a free scalar field
in flat space, with a kinetic term that is fourth order in derivatives. 

The classical Einstein-Hilbert action for a conformally flat
metric $g_{ab} = e^{2 \sigma}\eta_{ab}$ is
\be
S_{_{EH}}[g= e^{2\sigma}\eta]= \frac{1}{8\pi G}\ \int d^4 x \ \left[ 3 e^{2 \sigma} 
(\partial_a \sigma)^2 - \Lambda e^{4 \sigma} \right]\,,
\label{cEin}
\ee
which has derivative and exponential self-interactions in $\sigma$.
It is remarkable that these complicated interactions can be treated 
systematically using the the fourth order kinetic term of (\ref{flata}). 
These interaction terms are renormalizable and their anomalous 
scaling dimensions due to the fluctuations of $\sigma$ can be computed 
in closed form\cite{NJP,AntMot}. Direct calculation of the conformal 
weight of the Einstein curvature term shows that it acquires an 
anomalous dimension $\beta_2$ given by the quadratic relation,
\be
\beta_2 = 2 + \frac{\beta_2^2}{2Q^2}\,.
\label{betE}
\ee
In the limit $Q^2 \rightarrow \infty$ the fluctuations of $\sigma$ are suppressed 
and we recover the classical scale dimension of the coupling $G^{-1}$ with mass
dimension $2$. Likewise the cosmological term in (\ref{cEin}) corresponding 
to the four-volume acquires an anomalous dimension given by 
\be
\beta_0 = 4 + \frac{\beta_0^2}{2Q^2}\,.
\label{betL}
\ee
Again as $Q^2 \rightarrow \infty$ the effect of the fluctuations of
the conformal factor are suppressed and we recover the classical scale
dimension of $\Lambda/G$, namely $4$. The solution of the quadratic
relations (\ref{betE}) and (\ref{betL}) determine the scaling dimensions 
of these couplings at the conformal fixed point at other values of $Q^2$. 

The positive corrections of order $1/Q^2$ (for $Q^2 > 0$)
in (\ref{betE}) and (\ref{betL}) show that both $G^{-1}$ and
$\Lambda/G$ {\it flow to zero} at very large distances. Because both 
of these couplings are separately dimensionful, at a conformal fixed
point one should properly speak only of the dimensionless
combination $\hbar G \Lambda/ c^3 = \lambda$. By normalizing
to a fixed four volume $V= \int d^4 x$ one can show that the finite
volume renormalization of $\lambda$ is controlled by the anomalous 
dimension,
\be
2 \delta - 1 \equiv 2\, \frac{\beta_2}{\beta_0} - 1 = 
\frac{\sqrt{1 - \frac{8}{Q^2}} - \sqrt{1 - \frac{4}{Q^2}}}
{1 + \sqrt{1 - \frac{4}{Q^2}}} \le 0\,.
\label{scal}
\ee
This is the anomalous dimension that enters the infrared
renormalization group volume scaling relation,\cite{NJP}
\be
V \frac{d}{d V} \lambda = 4\, (2 \delta - 1)\, \lambda\,.
\label{renL}
\ee
The anomalous scaling dimension (\ref{scal}) is negative for all 
$Q^2 \ge 8$. This implies that the dimensionless cosmological term 
$\lambda$ has an infrared fixed point at zero as $V\rightarrow \infty$. Thus the 
cosmological term is {\it dynamically driven to zero} as $V\rightarrow \infty$
by infrared fluctuations of the conformal part of the metric described 
by (\ref{flata}).

No fine tuning is involved here and no
free parameters enter except $Q^2$, which is determined by the trace
anomaly coefficient $b'$ by (\ref{Qdef}). Once $Q^2$ is assumed
to be positive, then $2 \delta - 1$ is negative, and $\lambda$ is
driven to zero at large distances by the conformal fluctuations 
of the metric, with no additional assumptions. Thus the fluctuatiuons
of the conformal scalar degrees of freedom of the anomaly generated
effective action $S_{anom}$ may be responsible for the observed
small value of the cosmological dark energy density (\ref{cosmeas})
inferred from the supernova data. Note also that the fields $\varphi$ and $\psi$ 
are scalar degrees of freedom in cosmology which arise naturally from the 
effective action of the trace anomaly in the Standard Model, without the 
{\it ad hoc} introduction of an inflaton field. Recent progress in evaluating
their effects of these anomaly scalars in de Sitter space indicate that
they have potentially large effects at the cosmological horizon.\cite{DSAnom} 
Even in the absence of a complete theory of dynamical cosmological 
vacuum energy, it is reasonable to assume that the conformal fluctuations of 
$S_{anom}$ could be observable in the signatures of conformal invariance 
should be imprinted on the spectrum and statistics of the CMB.

\section{Conformal Invariance and the CMB}

Our earlier studies of fluctuations in de Sitter space suggest that the fluctuations 
responsible for the screening of $\lambda$ take place at the horizon scale.\cite{NJP} 
In that case then the microwave photons in the CMB reaching us from their surface 
of last scattering should retain some imprint of the effects of these fluctuations. 
It then becomes natural to extend the classical notion of scale invariant cosmological 
perturbations to full conformal invariance. In that case the classical spectral index 
of the perturbations should receive corrections due to the anomalous 
scaling dimensions at the conformal phase.\cite{sky} In addition to 
the spectrum, the statistics of the CMB should reflect the non-Gaussian 
correlations characteristic of conformal invariance. 

Consider first the two-point function of any observable ${\cal O}_{\Delta}$
with dimension $\Delta$. Conformal invariance requires 
\be
\langle{\cal O}_{\Delta} (x_1) {\cal O}_{\Delta} (x_2)\rangle
\sim \vert x_1-x_2 \vert^{-2\Delta}
\label{ODel}
\ee
at equal times in three dimensional flat spatial coordinates. In
Fourier space this gives
\be
G_2(k) \equiv\langle\tilde{\cal O}_{\Delta} (k) \tilde 
{\cal O}_{\Delta} (-k)\rangle \sim \vert k \vert^{2\Delta - 3} \,.
\label{G2}
\ee
Thus, we define the spectral index of this observable by
\be
n \equiv 2 \Delta - 3\ .
\label{index}
\ee
In the case that the observable is the primordial density fluctuation 
$\delta\rho$, and in the classical limit where its anomalous dimension 
vanishes, $\Delta \rightarrow p =2$, we recover the 
Harrison-Zel'dovich spectral index of $n=1$.

In order to convert the power spectrum of primordial density 
fluctuations to the spectrum of fluctuations in the CMB at large angular 
separations we follow the standard treatment relating the
temperature deviation to the Newtonian gravitational potential 
$\varphi$ at the last scattering surface, $\frac{\delta T}{T} \sim \delta 
\varphi$, which is related to the density perturbation in turn by 
\be
\nabla^2 \delta\varphi = 4\pi G\, \delta\rho \ .
\label{lap}
\ee
Hence, in Fourier space, 
\be
\frac{\delta T}{T} \sim \delta \varphi \sim 
\frac{1}{k^2}\frac{\delta\rho}{\rho}\ ,
\ee
and the two-point function of CMB temperature fluctuations is
determined by the conformal dimension $\Delta$ to be
\begin{eqnarray}
&&C_2(\theta) \equiv \left\langle\frac{\delta T}{T}(\hat r_1)
\frac{\delta T}{T}(\hat  r_2)\right\rangle
\sim \nonumber\\
&&\int d^3 k\left(\frac{1}{k^2}\right)^2 G_2(k) e^{i k\cdot 
r_{12}}
\sim \Gamma (2-\Delta) (r_{12}^2)^{2 - \Delta}\ ,
\label{C2}
\eea
where $r_{12} \equiv (\hat r_1 - \hat r_2)r$
is the vector difference between the two positions from which the 
CMB photons originate. They are at equal distance $r$ from the observer 
by the assumption that the photons were emitted at the last scattering 
surface at equal cosmic time. Since $r_{12}^2 = 2 (1- \cos \theta)r^2$, 
we find then
\be
C_2(\theta) \sim \Gamma (2-\Delta) (1-\cos\theta)^{2 - \Delta}
\label{Ctheta} 
\ee
for arbitrary scaling dimension $\Delta$.  

Expanding the function $C_2(\theta)$ in multipole moments,
\be
C_2(\theta) = \frac{1}{4\pi} \sum_{\ell} (2\ell + 1)
c_{\ell}^{(2)}(\Delta) P_{\ell} (\cos \theta)\ ,
\label{c2m}
\ee
\be
c_{\ell}^{(2)}(\Delta) \sim \Gamma(2-\Delta) \sin\left[ \pi 
(2-\Delta)\right]
\frac{\Gamma (\ell - 2 + \Delta)}{\Gamma (\ell + 4 -\Delta)}\ ,
\ee
shows that the pole singularity at $\Delta =2$ appears only in the $\ell = 
0$ monopole moment. This singularity is just the reflection of the fact 
that the Laplacian in (\ref{lap}) cannot be inverted on constant functions, 
which should be excluded. Since the CMB  anisotropy is defined by removing 
the isotropic monopole moment (as well as the dipole moment), the $\ell =0$ 
term does not appear in the sum, and the higher moments of the anisotropic 
two-point correlation function are well-defined for $\Delta$ near $2$. 
Normalizing to the quadrupole moment $c_2^{(2)}(\Delta)$, we find
\be
c_{\ell}^{(2)}(\Delta) = c_2^{(2)}(\Delta) 
\frac{\Gamma (6 - \Delta)}{\Gamma (\Delta) } 
\frac{\Gamma (\ell - 2 + \Delta)}{\Gamma(\ell + 4 - \Delta)}\ ,
\ee
which is a standard result. Indeed, if $\Delta$ is replaced by 
$p = 2$ we obtain $\ell (\ell + 1)  c_{\ell}^{(2)}(p) = 6 c_2^{(2)} (p)$, 
which is the well-known  predicted behavior of the lower moments ($\ell \le 30 $) 
of the CMB anisotropy where the Sachs-Wolfe effect should dominate.

Turning now from the two-point function of CMB fluctuations to higher point 
correlators, we find a second characteristic prediction of conformal invariance, 
namely non-Gaussian statistics for the CMB. The first correlator sensitive to this 
departure from gaussian statistics is the three-point function of the observable 
${\cal O}_{\Delta}$, which takes the form \cite{sky}
\be
\langle{\cal O}_{\Delta} (x_1) {\cal O}_{\Delta} (x_2) 
{\cal O}_{\Delta} (x_3)\rangle
\sim |x_1-x_2|^{-\Delta} |x_2- x_3|^{-\Delta} |x_3 - x_1|^{-\Delta}\ ,
\ee 
or in Fourier space,\footnote{Note that (\ref{three}) corrects two minor typographical
errors in eq. (16) of Ref. \cite{sky}}
\bea
&&G_3 (k_1, k_2) \sim \int d^3 p\ |p|^{\Delta -3}\,  |p + 
k_1|^{\Delta -3}\, 
|p- k_2|^{\Delta -3}\, \sim 
\frac{\Gamma\left( 3 - {\textstyle\frac{3\Delta}{2}} \right)}
{\left[\Gamma\left(\frac{3-\Delta}{2}\right)\right]^3}\times\nonumber\\
&&\hspace{-7mm}\int_0^1 du\,\int_0^1 dv\, \frac{\left[u(1-u)v\right]^{\frac{1-\Delta}{2}} 
(1-v)^{-1 + \frac{\Delta}{2}}}{
\left[u(1-u)(1-v)k_{_1}^2 + v(1-u)k_{_2}^2 + uv(k_{_1} + 
k_{_2})^2\right]^{3-\frac{3\Delta}{2}}}\,.
\label{three}
\eea
This three-point function of primordial density fluctuations gives rise to
three-point correlations in the CMB by reasoning precisely analogous as that
leading from Eqns.~(\ref{G2}) to (\ref{C2}). That is,
\bea
&& C_3(\theta_{12}, \theta_{23}, \theta_{31}) 
\equiv \left\langle\frac{\delta T }{T}(\hat r_1)
\frac{\delta T}{T}(\hat  r_2)\frac{\delta T }{T}(\hat  
r_3)\right\rangle \nonumber \\
&& \qquad\sim \int \frac{d^3 k_1\,d^3k_2}{
k_1^2\, k_2^2\, (k_1 + k_2)^2}\ G_3 (k_1, k_2)\, e^{i k_1\cdot r_{13}} 
e^{ik_2\cdot r_{23}}
\label{C3}
\eea
where $r_{ij}\equiv ({\hat r}_i-{\hat r_j})r$ and 
$r_{ij}^2=2(1-\cos\theta_{ij}) r^2$. 

In the general case of three different angles, this expression for the non-Gaussian
three-point correlation function (\ref{C3}) is quite complicated, although it can be 
rewritten in parametric form analogous to (\ref{three}) to facilitate numerical evaluation. 
In the special case of equal angles, it follows from its global scaling properties
that the three-point correlator is 
\be
C_3(\theta)\sim (1-\cos\theta)^{\frac{3}{2}(2-\Delta)}\ .
\ee
Expanding the function $C_3(\theta)$ in multiple moments as in (\ref{c2m})
with coefficients $c_{\ell}^{(3)}$, and normalizing to the quadrupole moment,
we find
\be
c_{\ell}^{(3)}(\Delta) =c_{2}^{(3)}(\Delta)
\frac{\Gamma (4+\frac{3}{2}(2-\Delta))}{\Gamma (2-\frac{3}{2}(2-\Delta))}
\frac{\Gamma (\ell-\frac{3}{2}(2-\Delta))}{\Gamma(\ell+2+\frac{3}{2}(2-\Delta))}\ .
\label{cl3}
\ee
In the limit $\Delta \rightarrow 2$, we obtain $\ell(\ell+1)c_{\ell}^{(3)}=6c_2^{(3)}$, 
which is the same result as for the moments $c_{\ell}^{(2)}$ of the two-point correlator 
but with a different quadrupole amplitude. The value of this quadrupole normalization 
$c_2^{(3)}(\Delta)$ cannot be determined by conformal symmetry considerations alone,
and requires more detailed dynamical information about the origin of conformal
invariance in the spectrum.

For higher point correlations, conformal invariance does not 
determine the total angular dependence. Already the four-point 
function takes the form,
\be
\langle{\cal O}_{\Delta} (x_1) {\cal O}_{\Delta} (x_2) 
{\cal O}_{\Delta} (x_3) {\cal O}_{\Delta} (x_4)\rangle
\sim \frac{ A_4}{{\prod_{i<j} r_{ij}^{2\Delta/3}} }\ ,
\ee
where the amplitude $A_4$ is an arbitrary function of the two 
cross-ratios, 
$r_{13}^2 r_{24}^2/r_{12}^2 r_{34}^2$ and 
$r_{14}^2 r_{23}^2/r_{12}^2 r_{34}^2$.
Analogous expressions hold for higher $p$-point functions. 

An important point to emphasize is that all of these results depend upon the 
hypothesis of conformal invariance on the spatially homogeneous and 
isotropic flat spatial sections of  geometries. This is only one way in which
conformal invariance may be realized, for example, if the universe
went through a de Sitter like inflationary expansion. That this is
actually related to the geometric symmetries of de Sitter space
is shown next.

\section{Conformal Invariance as a Consequence of de Sitter Invariance}

In cosmology the line element of de Sitter space is usually expressed in the form
\be
ds^2 = - d\tau^2 + a^2(\tau)\, d\vec x\cdot d\vec x = - d\tau^2 + e^{2H\tau}\, (dx^2 + dy^2 + dz^2)
\label{flatRW}
\ee
with flat spatial sections, and the Hubble parameter $H = \sqrt{\Lambda/3}$.
This de Sitter geometry has an $SO(4,1)$ symmetry group with $10$
Killing generators satisfying
\be
\nabla_a \xi_b^{(\alpha)} + \nabla_b \xi_a^{(\alpha)} = 0\,,\qquad \alpha = 1,\dots,10\,,
\label{Kil}
\ee
which leave the de Sitter metric invariant. In coordinates (\ref{flatRW}), (\ref{Kil}) becomes
\bes
\bea
&&\partial_{\tau}\xi_{\tau}^{(\alpha)} = 0\,,\\
&&\partial_{\tau} \xi_i^{(\alpha)} + \partial_i\xi_{\tau}^{(\alpha)}- 2H \xi_i^{(\alpha)} = 0\,,\\
&&\partial_i\xi_j^{(\alpha)} + \partial_j\xi_i^{(\alpha)} - 2Ha^2 \delta_{ij} \xi_{\tau}^{(\alpha)} = 0\,.
\label{cKv}
\eea\label{KilldS}\ees
\noindent
For $\xi_{\tau} =0$ we have the three translations, $\alpha = T_j$,
\be
\xi_{\tau}^{(T_j)} = 0\,,\qquad \xi_i^{(T_j)} = a^2 \delta_i^{\ j}\,,\qquad j = 1, 2, 3\,,
\ee
and the three rotations, $\alpha =R_\ell$,
\be
\xi_{\tau}^{(R_\ell)} = 0\,,\qquad \xi_i^{(R_\ell)} = a^2 \epsilon_{i\ell m}x^m\,,\qquad \ell = 1, 2, 3\,.
\ee
This accounts for $6$ of the $10$ de Sitter isometries which are self-evident in
the spatially flat homogeneous and isotropic Robertson-Walker coordinates (\ref{flatRW})
with $\xi_{\tau} =0$. The $4$ additional solutions of (\ref{KilldS}) have
 $\xi_{\tau} \neq 0$. They are the three special conformal transformations
of $\Real^3$, $\alpha = C_n$,
\be
\xi_{\tau}^{(C_n)} = -2H x^n\,,\qquad \xi_i^{(C_n)} = H^2a^2( \delta_i^{\ n} \delta_{jk}x^jx^k 
- 2 \delta_{ij}x^jx^n)
- \delta_i^n\,,\qquad n = 1, 2, 3\,,
\label{confspec}
\ee
and the dilation, $\alpha = D$,
\be
\xi_{\tau}^{(D)} = 1\,,\qquad \xi_i^{(D)} = H a^2\,\delta_{ij} x^j\,.
\label{dilat}
\ee
This last dilational Killing vector is the infinitesimal form of the finite dilational symmetry,
\bes
\bea
&&\vec x \rightarrow \lambda \vec x\,,\\
&& a(\tau) \rightarrow \lambda^{-1} a(\tau)
\eea
\ees
of de Sitter space. The existence of this symmetry explains why Fourier modes of
different $|\vec k|$ leave the de Sitter horizon at a shifted RW time $\tau$, so in
an eternal de Sitter space, in which there is no preferred $\tau$, one expects
a scale invariant spectrum.

The existence of the three conformal modes of $\Real^3$ (\ref{confspec}) implies in 
addition that any $SO(4,1)$ de Sitter invariant correlation function must decompose
into representations of the conformal group of three dimensional flat space.
Fundamentally this is because the de Sitter group $SO(4,1)$ {\it is} the conformal
group of flat Euclidean $\Real^3$, as eqs. (\ref{Kil})-(\ref{dilat}) shows explicitly.
Moreover, because of the exponential expansion in de Sitter space, the 
decomposition into representations of the conformal group become simple at 
distances large compared to the horizon scale $1/H$ \cite{comm97}. Thus
if the universe went through an exponentially expanding de Sitter phase
for many e-foldings when the fluctuations responsible for the CMB were generated, 
then the CMB should exhibit full conformal invariance in addition to simple scale 
invariance.  Neither the form nor magnitude of the CMB power or bispectrum
depend upon an inflaton or ``slow-roll" parameters as in conventional 
scalar models of inflation. 

Another quite distinct possibility for realizing conformal invariance from
de Sitter space is if the fluctuations due to the anomaly scalars are generated 
in the vicinity of the cosmological horizon at $r=1/H$ in the {\it static} coordinates
of de Sitter space, {\it i.e.}.
\be
ds^2 = - (1-H^2 r^2) dt^2 + \frac{dr^2}{(1-H^2r^2)} + r^2 d \Omega^2\,.
\label{dSstat}
\ee
Conformal invariance on the sphere $r=1/H$ leads to a different characteristic form of
the non-Gaussian bispectrum and higher angular correlations. This form and additional
relevant details will be presented in a forthcoming article.\cite{AntMazMot11}

\end{document}